# Chemical Short-Range Ordering in a CrCoNi Medium-Entropy Alloy

H.W. Hsiao[1,2&], R. Feng[3], H. Ni[1,2], K. An[3], J.D. Poplawsky[4], P.K. Liaw[5],

and J.M. Zuo[1,2*]

[1]Department of Materials Science and Engineering, University of Illinois at Urbana-Champaign, 1304 W Green St, Urbana, IL 61801, United States
[2]Fredrick Seitz Materials Research Laboratory, University of Illinois at Urbana-Champaign, 104 S Goodwin Ave, Urbana, IL 61801, United States
[3]Neutron Scattering Division, Oak Ridge National Laboratory, Oak Ridge, TN, 37831, United States
[4]Center for Nanophase Materials Sciences, Oak Ridge National Laboratory, Oak Ridge, TN, 37831, United States
[5]Department of Materials Science and Engineering, The University of Tennessee Knoxville, Knoxville, TN 37996, United States
&First two authors contributed equally to this work
*Corresponding Author jianzuo@illinois.edu

**Abstract:**

The exceptional mechanical strengths of medium/high-entropy alloys have been attributed to hardening in random solid solutions. Here, we evidence non-random chemical mixings in CrCoNi alloys, resulting from short-range ordering. A novel data-mining approach of electron-nanodiffraction patterns enabled the study, which is assisted by neutron scattering, atom probe tomography, and diffraction simulation using first-principles theory models. Results reveal two critical types of short-range-orders in nanoclusters that minimize the Cr–Cr nearest neighbors ($L1_1$) or segregate Cr on alternating close-packed planes ($L1_2$). The makeup of ordering-strengthened nanoclusters can be tuned by heat treatments to affect deformation mechanisms. These findings uncover a mixture of bonding preferences and their control at the nanoscopic scale in CrCoNi and provide general opportunities for an atomistic-structure study in concentrated alloys for the design of strong and ductile materials.

**Main**

High-entropy alloys (HEAs) belong to a new class of materials that are chemically concentrated in less explored phase spaces. Since their initial discovery[1,2], HEAs have attracted tremendous interest for their remarkable structural diversity[3-6] and unique electrical, magnetic, and mechanical properties[7,8], including high strength and great ductility[9]. While HEAs were initially modelled as random solid solutions[10], recent works have focused on the possibility of chemical short-range ordering (CSRO) and its unusual effects on microstructural, mechanical, and electronic properties[11-14]. Particular attention has been given to the medium entropy alloy (MEA) CrCoNi. This alloy, which belongs to a family of Cr-Mn-Fe-Co-Ni alloys[2], has demonstrated superior mechanical properties[15], irradiation resistance[15,16], and quantum criticality at low temperatures[17]. Especially, the close proximity of ferromagnetic (Co and Ni) and antiferromagnetic (Cr) interactions can lead to magnetic frustration, and the concept of magnetic-driven CSRO has inspired multiple experimental[18-20] and theoretical works[21-23].

The standard analysis of CSRO is by single-crystal diffraction[24]. Studies of binary alloys have demonstrated that CSRO produces characteristic diffuse scattering, and when measured by X-ray and neutron diffraction, a quantitative determination of atomic pair correlations can be made[24-26]. However, single-crystal-diffuse scattering analysis of MEA/HEAs has been performed only rarely and the interpretation of diffuse scattering from samples with different thermomechanical histories is extremely challenging[13]. Electrons can work with polycrystals. However, a significant challenge here is the lack of strong elemental contrast, which prevents direct imaging of CSRO at the atomic resolution. For the electron-diffraction study of the face-centered-cubic (fcc) CrCoNi, Zhang et al.[19] reported diffuse streaks along {111} in the energy-filtered [110] zone-axis-diffraction patterns (DPs), while Zhou et al. reported the $(\bar{3}11)/2$ diffuse spot in the [112] zone-axis[20]. However, it is far from being clear whether these observations support the theory-predicted CSRO[12,21,23]. In the absence of strong experimental evidence, it has been questioned whether CSRO in CrCoNi is a negligible effect[27].

Here we develop a data-driven electron-diffraction approach for the determination of local CSRO, using scanning electron nanodiffraction with energy filtering (EF-SEND), which overcomes the limitations of the traditional diffuse-scattering analysis with orders of magnitude improvement in the spatial resolution and electron-scattering power. The diffraction analysis is complemented by (1) small-angle neutron scattering (SANS) and atom probe tomography (APT) analyses of chemical fluctuations at a sub-nm scale and (2) diffraction simulations using first-principles theory models[23].

**Ordering in CrCoNi**

We prepared two CrCoNi samples from the same ingot, as described in Methods. Following the initial homogenization at 1,200 °C for 48 h, one sample was water-quenched (Sample WQ), while the other was heated treated (Sample HT) at 1,000 °C for additional 120 h aging, followed by furnace cooling. Both samples show the same fcc structure and similar mechanical response (*Suppl. Info. Figs. 1a,b*). The diffuse streaks along {111} directions reported by Zhang et al.[19] are observed in both samples here by EF-SAED (energy-filtered selected-area electron diffraction) (*Suppl. Info. Figs. 1c,e*). Atomic-resolution Z-contrast images obtained by scanning transmission electron microscopy (STEM) show relatively uniform contrast for both samples with no evidence of CSRO in the images (see insets in *Suppl. Info. Figs. 1d,f*); this finding is consistent with the previous report by Sales et al.[7].

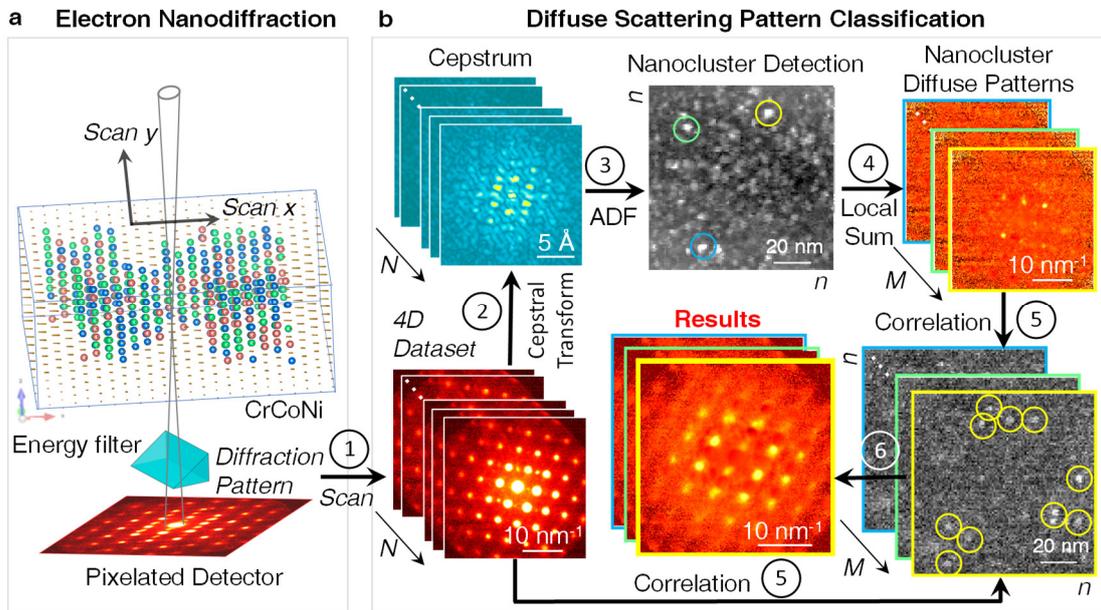

*Fig. 1. Data mining of nanodiffraction patterns for the detection of CSRO in CrCoNi. Six steps (1 to 6) are involved. (a) Step 1, energy-filtered electron diffraction patterns (DPs) are recorded using an ~ 1 nm sized probe for each probe position, with a pixelated detector. (b) Steps 2 to 6, the DPs are classified, based on diffuse scattering, and the CSRO diffuse patterns (DFPs) are identified (Steps 2 to 4), and then employed to map the CSRO-strengthened nanoclusters (Step 5). The final output is the measured DFPs (Step 6) from the same CSRO-strengthened nanoclusters located in Step 6. N is the number of collected DPs ($10^4$), M is the number of distinct DFPs ($10^1$), and n ($n^2 = N$) is the number of scan points. ADF stands for annular dark field imaging, using the cepstral signals here. Step 4 has three parts: one is to locate the nanoclusters and measure their DFPs by local summation, the second is to classify the distinct DFPs, and the third is to identify these formed by CSRO. Step 5, the correlation image between each distinct DFP and the background-subtracted DPs in the 4Ddataset is calculated to locate the nanoclusters with*

*the same CSRO.*

The seeming sameness of two CrCoNi samples from the above results demonstrates a major conundrum in the MEA/HEA research, namely, the lack of suitable probes for the chemical subtlety in these materials[3]. To overcome this challenge, we developed a nanodiffraction data-mining approach for the detection of CSRO (**Figure 1** and *Suppl. Notes 1*). First, a small convergent-angle (1.1 mrad) beam is formed for coherent nanodiffraction. Second, an energy filter removes the inelastic scattering background for diffraction pattern recording. Third, a stack of DPs is collected in a two-dimensional (2D) scan for a hyperspectral, four-dimensional (4D), dataset using a pixelated detector (**Figure 1a**). Data mining of the 4D dataset then allows for diffraction imaging. This powerful approach, known as 4D-STEM, has been demonstrated for the mapping of electrical, magnetic, and strain fields[28,29,30]. The data-mining process in **Figure 1b** extends 4D-STEM to diffuse-scattering analysis. Key to the analysis is the cepstral STEM-based imaging of nanoclusters (NCs) giving strong diffuse scattering[31]. Their detection then allows an identification of diffuse patterns for NCs. The DP correlation analysis and the resulted correlation images are subsequently used to determine the distribution of specific types of diffuse patterns. The sensitivity to weak diffuse scattering is improved as the amount of electron exposure is multiplied by the number of scan points, $N \sim 10^4$, while the spatial resolution is simultaneously improved from hundreds nm in EF-SAED to 1 nm in EF-SEND.

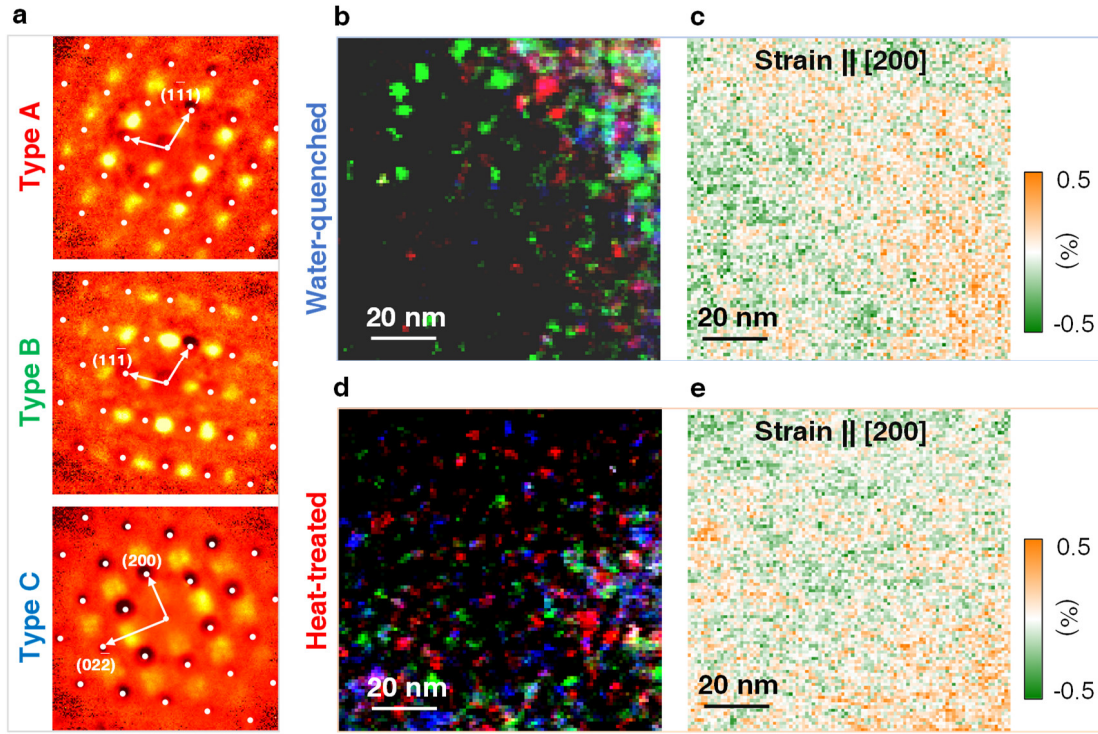

*Fig. 2 The identified CSRO diffuse scattering patterns and their distribution in CrCoNi. a) Three types (A, B, and C) of diffuse-scattering patterns along [011], detected by EF-SEND in both water-quenched (WQ) and heat-treated (HT) samples (the white dots mark Bragg peak positions). b) and d) are composite color images formed by integrating the diffuse-scattering intensity for Type A (red), Type-B (green), and Type-C (blue) for the WQ and HT samples, respectively. c) and e) show the measured strains along the [200] direction from Bragg peaks recorded by EF-SEND for the two samples, respectively.*

**Figure 2a** shows three types (A, B, and C) of diffuse-scattering patterns that were identified along the [011] zone axis from $10^4$ DPs collected for the two samples. The Bragg diffraction peaks and the spatially-invariant diffuse background in these patterns were removed, using the method described by Shao et al.[31] (*Suppl. Info. Note 1*). Among the three patterns, Types A and B are two variants of the same type with diffuse-intensity peaks at the special points of {111}/2, while in Type C, the broad diffuse peaks are observed at the special points of (100) and (110). These special points of diffuse scattering were previously observed in CuPt and CuPd, with the $L1_1$- and $L1_2$-type order, respectively[32]. Notably, the $L1_2$-type diffuse peaks are split in **Fig. 2c**, and the same feature is also present in CuPd[32].

**Figure 2b,d** show two color composites of three correlational images obtained by matching DPs in the hyperspectral dataset with the Type-A, B, and C diffuse patterns, respectively for Samples WQ and HT. Both composites show that the observed diffuse scattering came from local NCs. In Sample WQ, the NCs are dominantly $L1_1$-type of ~6 nm in size (red and green), while in Sample HT, $L1_1$-type NCs are smaller at ~2 nm (red and green) and $L1_2$-type NCs (blue) become more prevalent. The distribution of CSRO NCs also changes, from being heterogeneous in Sample WQ to comparatively homogeneous in Sample HT. This difference is also seen in the strain maps obtained from the Bragg diffraction peak analysis[33] of the same hyperspectral datasets (**Figs. 2c,e** and *Suppl. Info. Note 2*). The observed strain homogenization correlates with the reduction of $L1_1$-type NCs, which suggests that these NCs are likely lattice distorted.

**The atomic structure models of CSRO** CSRO in CrCoNi has been examined theoretically by Tamm et al.[21] and Ding et al.[12], using the Monte Carlo method combined with density functional theory (DFT) calculations. Their calculations, which were summarized in the Warren-Cowley (WC) short-range ordering parameters ($\alpha_{ij}$, with *i* and *j* for each element)[24], gives the negative $\alpha_{CrNi}$ and $\alpha_{CrCo}$ and positive $\alpha_{CrCr}$ for a preference of additional Cr–Co and Cr–Ni nearest neighbors at the expense of the Cr–Cr pairs. The origin of CSRO in CrCoNi is magnetic[21-23]. For example, in CrCoNiFe, the anti-ferromagnetic Cr is surrounded by the ferromagnetic Ni, Fe, and Co in an alloyed $L1_2$ structure driven by a reduction in energy[21,22].

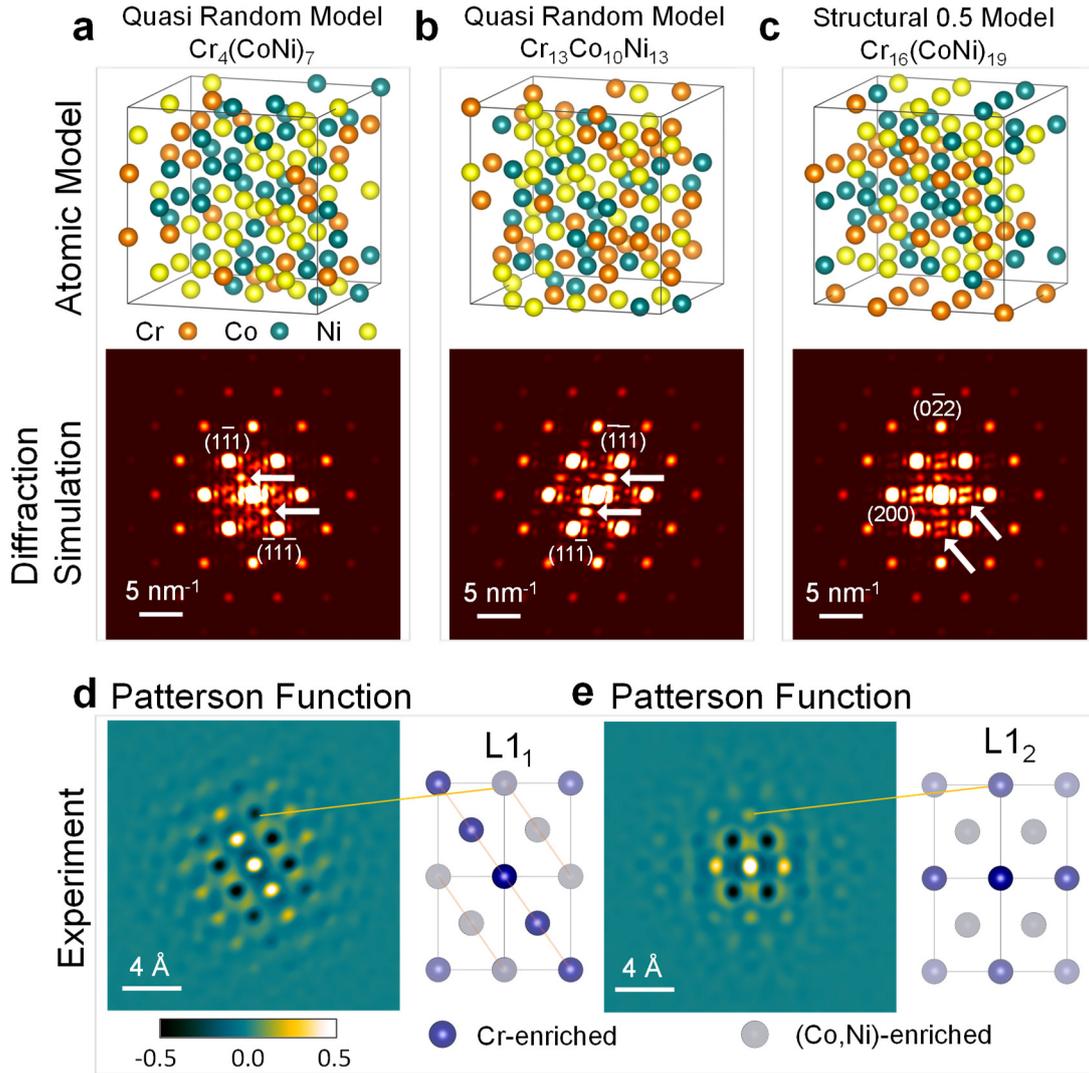

*Fig. 3 Short-range ordering in CrCoNi. Selected atomic structural models from first-principles calculations with (a) and (b) from quasi-random configurations and (c) from CSRO configuration (Structural 0.5 model)[23] and their corresponding simulated diffraction patterns. (d) and (e) are experimental Pattern functions and their structural model.*

To collate with the theoretical predictions, we examined the library of atomic-structure models built by Walsh et al.[23] by diffraction simulations (*Suppl. Info. Note 3*). The models were built with different degrees of CSRO and spin ordering, covering a range of stoichiometry[23]. **Figures 3a,b,c** present three models that yield diffuse peaks at the special Brillouin zone points. The models in **Figs. 3a,b** belong to the simulated quasi-random configurations with $\alpha_{ij} = 0$, while the model in **Fig. 3c** belongs to the structural 0.5 model, as simulated with $\alpha_{CrCr} = 0.5$ and $\alpha_{CrCo} = \alpha_{CrNi} = -0.25$[23]. The diffuse intensity tends to be stronger in the off-stoichiometric models, i.e., the models in **Figs. 3a,b,c** are all of off-stoichiometry. They are part of the trends that we found with the $L1_1$-type features occurring most frequently in off-stoichiometry quasi-random models and the $L1_2$-type most frequently observed in the off-stoichiometry structural models.

Next, we examine the atomistic origin of the $L1_1$- and $L1_2$-type diffuse scattering, using the Patterson-functions obtained from diffuse patterns in **Figs. 2a,c**. The negative nearest-neighbor Patterson peaks in **Fig. 3e** indicate the $L1_2$-type ordering. In contrast, Patterson function peaks in the same {111} plane have the same sign in the $L1_1$-type ordering, while the signs are opposite between neighboring planes (**Fig. 3d**). The Patterson-function peak for the site $i$, $P_{0i}$, in binary alloys relates to the WC parameter, according to $P_{0i} \propto m_0 m_i \alpha_{0i}$, with $i$ indexes the atomic site, and $m$ stands for the elemental composition[24]. In a ternary alloy, $P_{0i}$ is weighted by the difference in the atomic-scattering potential, and for CrCoNi, the largest difference is between Cr and (Co, Ni) over the experimental range of scattering angles (*Suppl. Info. Fig. S2*). Based on this trend, we conclude that the negative $P_{01}$ for the nearest neighbors is associated with the formation of unlike pairs of Cr-Co and Cr-Ni in the $L1_2$-type ordering. The estimated $\alpha_{01}$ is at $-0.18$ for the $L1_2$ CSRO, which is consistent with the Monte-Carlo DFT theory predictions[12,21] (Details in *Suppl. Note 4*). The $L1_1$-type ordering, on the other hand, goes beyond the nearest neighbors to the distance of 4.4 Å, and it is formed by Cr and (Co,Ni) clustering on alternating {111} planes.

Local chemical clustering contributes to chemical fluctuations in the samples. One indicator of such fluctuations is the observed large misfit volume in CrCoNi[23,27]. However, quantification of local chemical fluctuations has been difficult. We performed chemical analyses on the CrCoNi samples, using multiple probes. First, STEM energy-dispersive X-ray (EDX) chemical mapping of regions studied by diffraction at 1 nm spatial resolution shows no evidence of decomposition. Second, SANS detects additional scattering objects from the matrix solid solution with the Guinier radius of the gyration, $R_g$ ~ 3.0 Å, in both Samples WQ and HT (*Suppl. Info. Fig. S3 and Suppl. Note 5*). Third, the reconstructed APT volumes of samples HT and WQ show no signs of order or decomposition (*Suppl. Info. Fig. 4 and Note 6*). A closer inspection of APT data, based on the partial radial distribution function (RDF) analysis (*Suppl. Info. Note 7*), using Ni and Co atoms, clearly shows the evidence for Ni and Co clustering extending to ~1 nm radius. The negative Ni-Cr and Co-Cr correlations indicate Cr depletion in Ni- or Co-rich clusters, while Ni-Co correlations are largely absent (*Suppl. Info. Figs. S4b,e*). The partial RDFs also indicate small chemical fluctuations in Sample WQ with the deviations from the bulk normalized composition observed in the APT volume.

**Impacts of CSRO on dislocation slips**

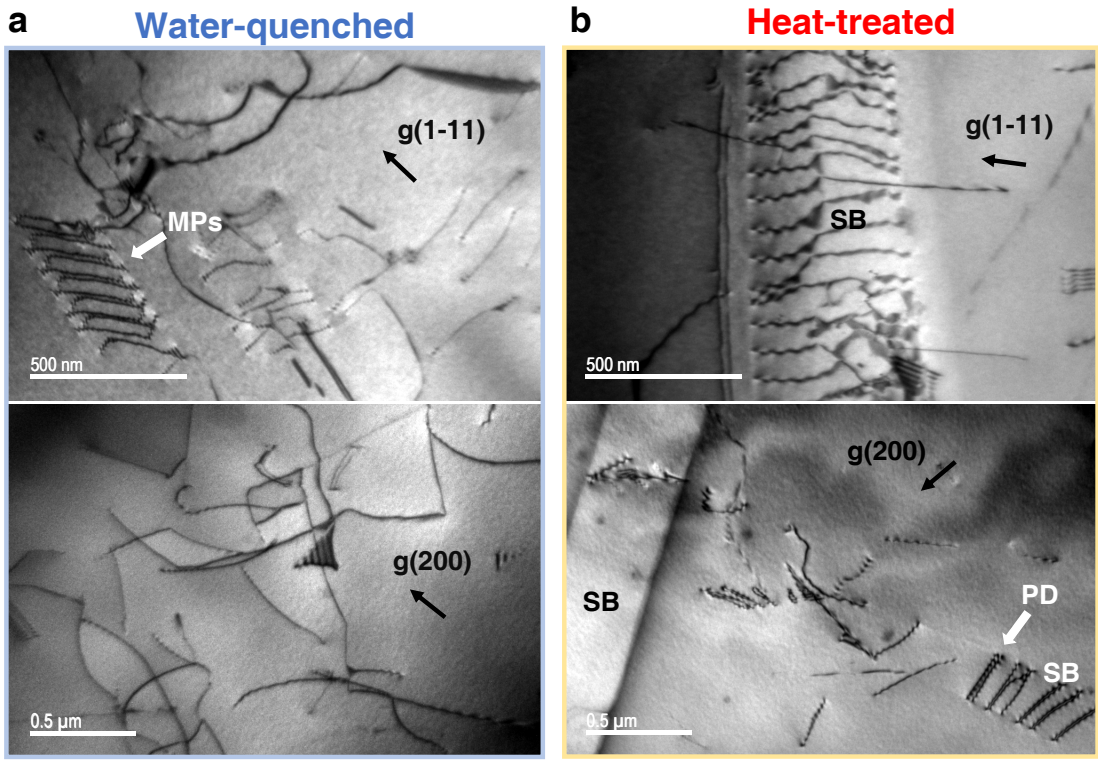

*Fig. 4 **Dislocation TEM micrographs show different behaviors in the slightly deformed crystals before and after additional thermal annealing.** (a) Two-beam bright-field images of Sample WQ shows long and curved dislocations. Multi-poles (MPs) are also observed here. **(b)** Dislocations in Sample HT form extensive slip bands (SBs), where paired dislocations (PDs) are commonly seen in the SBs.*

Experimentally, we find the predominance of different types of CSRO can have a pronounced impact on dislocation slips. **Figure 4** summarizes our TEM observations along two orientations. In Sample WQ, wavy dislocations were seen to be mixed with the localized planar slip, which changes drastically to extensive planar slip in Sample HT, with the latter observation agreeing with Zhang at el.[19]. In Sample WQ, localized planar slips sometimes form dislocation multipoles (MPs) (**Fig. 4a** and Suppl. Note 8). These features can be attributed to the hardening effects of the finely dispersed $L1_1$-type ordered NCs[34]. In Sample HT, the leading dislocations in the planar slip tend to form paired dislocations (PDs) at a short spacing, compared to other dislocations in the slip band (**Fig. 4b**). The formation of PDs is due to the destruction of the $L1_2$-type CSRO by the leading dislocation, which creates higher energy at the diffuse antiphase boundary, as the second dislocation partially heals the diffuse antiphase boundary created by the lead dislocation and thus, reduces the energy cost for dislocation motion[35,36]. The destruction of CSRO introduces the so-called glide plane softening effect[34] for the formation of the pronounced planar slip, as seen in **Fig. 4b**.

**Additional discussions and conclusion**

The discovery of the theory-predicted $L1_1$-type CSRO supports the significant role of magnetic interactions in CrCoNi. While the importance of magnetism in the development of alloys' microstructures and deformation mechanisms has been known for many decades[37], the atomistic mechanisms have only been addressed recently, largely through computations[12,21,23,38].

The $L1_1$ ordered nanoclusters are more dominant than the $L1_2$ ones in the rapidly cooled sample. The diffuse peaks at $\{111\}/2$ positions (**Fig. 2a**), which characterize the $L1_1$-type ordering, are reproduced by diffraction simulations in some of the off-stoichiometric, quasi-random, configurations simulated by Walsh et al.[23]. These configurations are either Cr or Co poor, and their formation is likely associated with the elemental diffusion of Cr or Co driven by the magnetic interactions. This scenario is consistent with the chemical heterogeneity observed by SANS and APT. Interestingly, the yield strengths of the two differently prepared samples are similar (*Suppl. Info Fig. S1*), which we attribute to the presence of $L1_1$-type and $L1_2$-type CSRO in both cases. The promotion of the $L1_2$-type CSRO is associated with the change from the predominant wavy dislocation slip in Sample WQ to planar slip in Sample HT.

Lastly, we note that neither the $L1_1$ nor $L1_2$ CSRO models produce the $(\bar{3}11)/2$ diffuse spot observed by Zhou et al.[20] in their cold-rolled CrCoNi plate sample[39]. This trend, together with our observations of CSRO-strengthened nanoclusters in the matrix of randomly mixed solid solutions, demonstrate experimentally the multiplicity of local bonding preferences in CrCoNi, dependent on the thermal and mechanical history. These discoveries are made possible with the establishment of a novel and sensitive diffraction characterization method, which overcomes the limitations of previous approaches based on single-crystal diffraction[40] or electron diffraction of selected volumes[19,20,32].

In conclusion, we have found the $L1_1$ and $L1_2$ types CSRO in the WQ and HT CrCoNi, respectively. The direct measurement and analysis of electron diffuse scattering unambiguously settle the prior question of whether CSRO exists in this alloy. The observation of maximum diffuse peaks at or near special points inside the Brillouin zone indicates a commonality with CSRO in binary alloys from the contribution of valence electrons, but with a special twist of strong magnetic interactions in CrCoNi. The significance of these confluent factors is highlighted by the dramatical alternation of dislocation-slip behavior with the promotion of $L1_2$ CSRO. Future theoretical studies are required to verify the nature of the $L1_1$ CSRO, as well as its contribution to hardening.

**Acknowledgments:** HWH and JMZ are supported by Intel and a SRI grant from GCOE, UIUC and NSF DMR-1828671. R. Feng thanks for the support from the Materials and Engineering Initiative at the Spallation Neutron Source (SNS), Oak Ridge National

Laboratory (ORNL). PKL is supported by the National Science Foundation (DMR-1611180 and 1809640) and the US Army Research Office (W911NF-13–1-0438 and W911NF-19–2-0049). A portion of this research used resources at the High Flux Isotope Reactor (HFIR) and SNS, a U.S. Department of Energy (DOE) Office of Science User Facility operated by the ORNL. APT was conducted at the ORNL's Center for Nanophase Materials Sciences (CNMS), which is a U.S. DOE Office of Science User Facility. The authors would like to thank Dr. Ken Littrell and Dr. Dunji Yu for their help in neutron-scattering measurements. The authors also thank James Burns for his assistance in performing APT experiments.

**Author contributions:** HWH, RF, and HN contributed to the experimental data collection, analysis, and their description. KA provided the support and guidance on the neutron scattering. JDP performed the APT experiments and analyzed the data. PKL provided the HEA sample, and comments and discussions on the paper. JMZ directed the research, developed the data-mining approach, contributed to data analysis, and wrote the manuscript with help of all authors.

**Competing interests**: The authors declare no competing interests.

**Materials & Correspondence:** All correspondence and material requests should be addressed to Jian-Min Zuo, jianzuo@illinois.edu.

**METHODS**

*Sample preparation and materials characterization:* The CrCoNi alloys were prepared from high purity (> 99.9% weight percent) Cr, Co, and Ni by arc melting and then drop casting under an argon atmosphere. To ensure the compositional homogeneity, the melting and casting processes were repeated five times. The prepared alloys were then cut into two pieces and homogenized at 1,200 °C for 48 h. After that, one was immediately cooled by water quenching, and another was aged at 1,000 °C for 120 h, followed by furnace cooling to room temperature. To avoid the oxidation effect, both samples were sealed in a vacuumed quartz capsule. The differently prepared samples were then fabricated into tensile bars for the tensile mechanical testing.

The monotonic tension was performed on an MTS Model 810 servohydrauli machine at room temperature with a strain rate of $1 \times 10^{-3}$ s$^{-1}$. The gauge section of the tensile samples has a dimension of 25.4 × 3 × 2 mm. Sample surfaces were mechanically polished to 1,200 grits before the tests. The tensile strains were measured by an MTS extensometer. Three tensile specimens were tested to ensure the reliability of the results. Neutron diffraction was performed on the VULCAN Engineering Materials Diffractometer at the Spallation Neutron Source (SNS), Oak Ridge National Laboratory (ORNL)[41,42]. Small-angle neutron scattering (SANS) was collected on the CG2 General-Purpose SANS at the High Flux Isotope Reactor (HFIR)[43]. Two-sample detector distances of 12 m and 1.5 m were used to measure the low Q ($4\pi sin\theta/\lambda$) range of 0.0061 to 0.117 Å$^{-1}$ and the high Q range of 0.044 to 0.722 Å$^{-1}$, respectively.

The APT specimens were fabricated, using the standard lift out and sharpening methods, as described by Thompson *et al.*[44]. Wedges were lifted out, mounted on Si microtip array posts, sharpened, using a 30 kV Ga$^+$ ion beam, and cleaned, employing a 2 kV Ga$^+$ ion beam. The APT experiment was run, utilizing a CAMECA LEAP 4000XHR in a voltage mode with a 50 K base temperature, 20% pulse fraction, a 0.5% detection rate, and a pulse repetition rate allowing for all elements to be detected. The APT results were reconstructed and analyzed, using the CAMECA's interactive visualization and analysis software (IVAS 3.8).

***Transmission electron microscopy:*** The TEM samples were prepared from the end of the tested tensile bars by mechanical cutting and polishing. The samples were further thinned down to electron transparency by twin-jet chemical etching, using an electrolyte consisting of 95% (volume percent) ethanol and 5% perchloric acid at a temperature of - 40 °C and an applied voltage of 30 V. TEM and STEM were carried, employing a JEOL 2010 TEM (JEOL USA, operated at 200 kV) and a Themis Z STEM (Thermo Fisher Scientific, USA, probe corrected and operated at 300 kV), respectively. Atomic-resolution images were collected, using a focused probe with a semi-convergence angle of 21.4 mrad and the detector inner cutoff angle of 40 mrad. For dislocation analysis, the samples were tilted to different two-beam diffraction conditions. The X-ray EDS area analysis was performed on thin sample areas of ~ 50 nm in thickness (identified by the convergent beam electron diffraction), using the Themis Z STEM.

*Energy-filtered scanning electron nanodiffraction*: Regions close to where EDS was conducted were selected for diffraction analyses. EF-SEND was performed, utilizing the Themis Z STEM in the microprobe mode with a beam of 40 pA, 1.1 mrad semi-convergence angle, and 1.2 nm in full-width half-maximum. A drift correction algorithm was applied to reduce the sample drift effect. The DP datasets were collected in 100 x 100 pixel scans over sample areas of 100 x 100 $nm^2$. Energy filtering was achieved, employing a Gatan image filter (GIF) camera and the energy selection slit width of 10 eV. Supplemental Information Fig. S5 provides examples of the record DPs.

# Supplementary Materials for

## Chemical Short-Range Ordering in a CrCoNi Medium-Entropy Alloy

H.W. Hsiao[1,2]†, R. Feng[3]†, H. Ni[1,2], K. An[3], J.D. Poplawsky[4], P.K. Liaw[5] and J.M. Zuo[1,2]*

Correspondence to: jianzuo@illinois.edu

**This file includes:**

Supplementary Text
Figs. S1 to S15
References

**Supplementary Notes**

Note 1. Data mining of four-dimensional diffraction datasets

To identify characteristic local diffuse scattering patterns, we developed the data mining approach outlined in **Fig. 1** in the main text. Here we provide further details on the major steps in our data mining approach.

*A. Bragg reflections removal and cepstral analysis*

The success of data mining is greatly helped by removing the strong transmitted beam and Bragg diffraction spots in the recorded DPs. **Fig. S6** demonstrates this improvement using the virtual annular dark-field (ADF) images reconstructed from the collected 4D-DD of the water-quenched sample (Sample WQ) as example. The virtual ADF image of total scattering is obtained by integrating diffraction intensity in as-recorded DPs with the inner and outer cutoff angles of 3.2 mrad and 32.1 mrad, respectively. Both Bragg reflections and diffuse scattering are included here. The ADF image reveals nano-clusters (NCs) of few nanometers in size with weak bright contrast on a non-uniform background (**Fig. S6A**). The background ADF intensity variation across the scanned area suggests changes of electron diffraction conditions, e.g., in the crystal thickness and orientation. The ADF image constructed from diffuse scattering only (**Fig. S6B**) shows improved contrast for the NCs. However, the same background variation in **Fig. S6A** is also observed here. **Fig. S6C** presents the cepstral ADF image, where NCs are clearly observed, including these that weakly appears in **Figs. S6A&B**. The contrast enhancement and the removal of the non-uniform background in **Fig. S6C** greatly assist the detection of NCs.

We use the log method to remove the Bragg reflections and the background intensity in a region averaged DP (the reference DP) using following equation described in ref (*1*):

$$\Delta I(\vec{k}) = log[I(\vec{k})] - log[I_o(\vec{k})] = log\left[\frac{I(\vec{k})}{I_o(\vec{k})}\right] \quad (1)$$

where $\vec{k}$ is the scattered wave vector and $I_0(\vec{k})$ represents the intensity in the reference DP $I_o(\vec{k})$, while $I(\vec{k})$ is the intensity in a single diffraction pattern in the 4D-DD. The diffraction pattern, $I(\vec{k})$, can be separated into two parts under the approximations that electron diffuse scattering, $I_D(\vec{k})$, is weak as shown in **Fig. S5B**, and the wave function associated with diffuse scattering has a random phase, hence

$$I(\vec{k}) = I_o(\vec{k}) + \bar{I}_o(\vec{k}) * I_D(\vec{k}) \quad (2)$$

where $\bar{I}_o(\vec{k})$ is the thickness averaged reference DP. The convolution and thickness averaging reflect that there are many beams in $I_0(\vec{k})$, each contributing to the total electron diffuse scattering as the beam travels through the entire sample thickness (*2, 3*). Using

$$log[I(\vec{k})] \approx log\, I_o(\vec{k}) + \Lambda(\vec{k}) * I_D(\vec{k}) \quad (3)$$

where $\Lambda(\vec{k}) = \bar{I}_o(\vec{k})/I_o(\vec{k})$, we have

$$\Delta I(\vec{k}) \approx \Lambda(\vec{k}) * I_D(\vec{k}) \quad (4)$$

Thus, the difference log intensity obtained in equation (1) is simply the diffuse scattering intensity convoluted by $\Lambda(\vec{k})$, which in a thin sample with the strong transmitted beam and relative constant intensity in the beam is simply the Fourier transform (FT) of the beam shape function.

*B. Cepstral analysis*

Difference cepstrum ($dC_p$) analysis of electron diffuse scattering is performed according to

$$dC_p(\vec{r}_p) = FT\left\{log\left[\frac{I(\vec{k},\vec{r}_p)}{I_o(\vec{k})}\right]\right\} \quad (5)$$

at the probe position $\vec{r}_p$. The interpretation of $dC_p$ is made based on the separation of the fluctuating part of the scattering potential ($U_1$) from the average scattering potential ($U_o$). The scattering potential seen by a nanosized electron beam is thus

$$U(\vec{r},\vec{r}_p) = U_o(\vec{r}) + U_1(\vec{r},\vec{r}_p) \tag{6}$$

where $U_1(\vec{r},\vec{r}_p)$ varies with the electron probe position $\vec{r}_p$. Diffraction by $U(\vec{r},\vec{r}_p)$ gives the local diffraction pattern $I(\vec{k},\vec{r}_p)$, while diffraction by $U_o$ gives $I_0(\vec{k})$. For example, a random alloy would yield $\langle U_1(\vec{r},\vec{r}_p)\rangle_{\vec{r}_p} = 0$. According to Shao et al. (*4*), $dC_p$ gives the Patterson function of the fluctuating scattering potential multiplied by a shape function, which is the FT of a top-hat function in a diffraction pattern where the transmitted beam is much stronger than the diffracted beams.

**Figure S7** shows an example of Bragg reflection removal and difference cepstrum analysis for two local areas, 1 and 2, as marked in **Fig. 6C**. These two areas show strong and weak contrast in the cepstral ADF image. The log difference DPs shown in **Fig. S7A,B** are obtained by averaging $log\left[\frac{I(\vec{k},\vec{r}_p)}{I_0(\vec{k})}\right]$ over the marked box regions, respectively. As these examples show, the log method effectively removes the strong Bragg reflections in the recorded DPs, while the cepstrum detects harmonic signals in the diffuse scattering.

*C. Identification of diffuse diffraction patterns*

A peak finding algorithm is used to locate NCs above an intensity threshold in the cepstral ADF image. Diffuse scattering patterns within each NC are averaged to obtain the cluster averaged DPs $[\overline{\Delta I_i}(\vec{k}), i = 1 \ldots n_c]$. These patterns can come from surface oxides, secondary phases, as well as CSRO. A majority of NCs seen in **Fig. 6C** are surface oxides. These nano-sized oxides are too small to be seen in selected area electron diffraction (SAED, **Fig. S1C&E**), but because of the enhanced diffraction sensitivity in EF-SEND, they are detected together with CSRO.

*D. Surface Oxides*

Surface oxides formed by oxidation also contribute to diffuse scattering captured by EF-SEND. To identify diffuse scattering DPs belonging to oxides, we performed EF-SEND on the samples that were exposed to air after 30 days using a Hitachi H9500 TEM equipped with a Gatan imaging filter (GIF). The DP shows continuous rings in this case, which are indexed using the NiO {111},{200} and {220} reflections. These rings were not initially observed in our freshly prepared TEM samples, as demonstrated by EF-SAED in **Fig. S1C&E** and the average SEND pattern in **Fig. S5A**. **Figure S8** identifies the surface oxide nano clusters. Using the indexed oxide reflection rings in **Fig. S8A**, we then identified the surface oxides among the cluster averaged DPs. The four DPs in **Fig. S8C** are cluster averaged DPs for NCs marked from 1 to 4 in **Fig. S8B.** This result demonstrates that the often-used NC imaging using diffuse intensity by SAED alone is not sufficient for the detection of CSRO. It is critical to exclude the surface oxide contribution from diffuse scattering analysis, which is only possible by spatially separating the oxide nanoparticles and the CSRO NCs in the collected 4D-DDs.

*E. Correlation imaging and the measurement of electron diffuse scattering*

To measure electron diffuse scattering from CSRO NCs, we follow the procedure illustrated in **Fig. 1** of the main text. First, three diffraction templates belonging to CSRO are created from the cluster averaged log difference DPs. These templates are then used to obtain the correlation images, using the normalized cross correlation coefficient (NCC)

$$\gamma_A = \frac{\sum_{x,y}\{[I_A(x,y)-\overline{I_A}]\cdot[I_B(x,y)-\overline{I_B}]\}}{\sqrt{\{\sum_{x,y}[I_A(x,y)-\overline{I_A}]^2\}\cdot\{\sum_{x,y}[I_B(x,y)-\overline{I_B}]^2\}}} \tag{7}$$

where $I_A(x,y)$ and $I_B(x,y)$ are intensities of the pixel $(x,y)$ in the DP $A$ and $B$ (with Bragg reflection removed), respectively, and $\overline{I_A}$ and $\overline{I_B}$ are mean intensities of DP $A$ and $B$, respectively. DP $B$ is taken as the diffraction template, while DP $A$ is one of DP in the collected 4D-DD. The value of $\gamma_A$ is subsequently used to form the correlation image. The value of $\gamma_A$ ranges from -1 to 1 with $\gamma_A = 1$ indicating complete similarity. From the correlation image we then identify NCs having $\gamma$ values to be equal to or greater than a fixed, pre-defined, threshold value. DPs from these selected NCs are subsequently averaged to generate new diffraction templates. This process is then repeated to refine the diffraction templates. The principle of using NCC to define similarity among DPs is similar to DP clustering employing the k-means method (5), which is broadly defined as the unsupervised classification of patterns.

Note 2. Strain mapping

Strain analysis via SEND is based on Bragg's law by measuring the change in distance between diffracted beams. The beam in the diffraction pattern appears as a disk because of the beam's semi-convergence angle (1.1 mrad for SEND). For each diffraction pattern, the positions of 9 selected beams, which include the center beam and 8 diffracted beams of low-order reflections, were determined using the circular Hough transform method described by Yuan et al. (6). The method works by applying first the Sobel filter to the SEND pattern to filter out the disk edge. Then, circular Hough transform is applied on the filtered SEND pattern, which transforms the filtered pattern of edge circles into a pattern of Hough transformation peaks, with each peak marking the position of a detected circle. The peak position was measured by fitting using a Lorentzian peak model. This measurement was applied to all 9 selected beams, and a 2D reciprocal lattice with $\vec{G_1}$ = (002) and $\vec{G_2}$ =(2-20) as the basis vectors was then determined from the measured disk positions. The calibration was performed using a standard Si specimen. **Figure S9** shows the obtained xx, yy and xy strain maps, plus the rotation maps, for Samples WQ and HT, respectively.

Note 3. Diffraction Simulation using first-principles theory models

The atomistic models published by Walsh et al. (*7*) were examined by diffraction simulation for comparison with our experimental DPs. All configurations considered here belong to the fcc crystal with 108 atoms in a volume with 3x3x3 unit cells. The configurations have a range of compositions from stoichiometric CrCoNi to off-stoichiometric compositions. The Warren-Cowley (WC) SRO parameters are used to generate the atomic configurations.

**Figure S10** summarizes the atomistic models we included in our diffraction simulation. For the Structure 0.5 model, the following WC SRO parameters are used: $\alpha_{CrCr} = 0.5$, $\alpha_{CrCo} = -0.25$, $\alpha_{CrNi} = -0.25$, $\alpha_{NiCo} = 0.25$, $\alpha_{CoCo} = 0.0$ and $\alpha_{NiNi} = 0.0$. And $\alpha_{ij} = 0$ is used for the quasi-random models. The WC SRO parameters for the Tamm and Ding models were obtained from the Monte Carlo optimization of on-lattice DFT simulations (*8, 9*). In the maximum spin (Max Spin) model, the placements of Co and Cr↑Cr↑ are optimized simultaneously, fully segregating these species.

Simulated DPs in **Fig. S10** are the sum of DPs obtained from different configurations in the same model. Diffuse intensity peaks near the special positions of (100) and (110) are observed in Tamm, Structure 0.5, Max Spin and spin ordered (not shown) models. The diffuse intensity is stronger in the off-stoichiometry Structure 0.5 model than in Tamm model, while the diffuse peak feature in Max Spin model is inconsistent with the experiment.

**Figure S11** examines the fluctuations within the off-stoichiometry structure 0.5 and quasi random atomistic models as seen by diffraction. Each configuration contains 3x3x3 unit cells and approximately 1 nm³ in volume.

Next, we performed diffraction simulations along the [112] zone axis orientation (**Fig. S12**). This zone axis was recently identified and employed for the study of CSRO in CrCoNi by Zhou et al. (*10*). A diffuse spot at $(\bar{3}11)/2$ was identified and attributed to CSRO. The question in hand is whether the CSRO predicted by the structure 0.5 model also produces this observation. The simulation results for the off-stoichiometry quasi-random and structure 0.5 models show that diffuse scattering occurred at (1-10) position in the structure 0.5 model, while no feature is observed in the quasi-random model. Neither of these models produce visible diffuse spot at $(\bar{3}11)/2$.

Note 4. Estimation of Warren-Cowley Short-Range Order Parameter:

SRO parameter can be determined from the Patterson function obtained from the recorded diffuse scattering patterns, as shown in **Figs. 3D,E** of the main text. Following Cowley (*11*), the SRO parameter $\alpha_{ij}$ relates the elemental composition $m_A$ and $m_B$, between site $i$ and $j$ by

$$\alpha_{ij} = 1 - \frac{P_{ij}^{AB}}{m_A m_B} \tag{8}$$

where $P_{ij}^{AB}$ is the probability of finding a B atom at site $j$ from an A atom at site $i$.

Under the kinematical diffraction condition and in absence of the size effect, the diffuse scattering intensity ($I_d$) of a disordered binary alloy is given by following as function of the reciprocal space coordinate $\vec{u}$ [19]

$$I_d(\vec{u}) = N m_A m_B (f_A - f_B)^2 \sum_i \alpha_{0i} \exp(2\pi i\, \vec{u} \cdot \vec{r}_i) \tag{9}$$

where the index $i$ is over the atomic sites $\vec{r}_i$. The Patterson function of the diffuse scattering $[P(\vec{r})]$ can be derived as

$$P(\vec{r}) = N m_A m_B \sum_i [\alpha_{0i}\{(\rho_A - \rho_B) * (\rho_A - \rho_B)\}] * \delta(\vec{r} - \vec{r}_i) \tag{10}$$

where the $\rho_A$ and $\rho_B$ represent the atomic potential of element A and B, respectively. The first term ($r_0 = 0$) in the equation (10) yields the zero-order peak at the center of the Patterson function map, and the second and high order terms correspond to the Patterson peaks for the first order (nearest) and higher order inter-distances between the projected 2D lattice sites. The scattering potential of transition metals $\rho$ is much more localized compared to the lattice site distances, so the scattering potential at the $j$-th site contributes very little to the $i$-th site. Therefore, the diffuse scattering Patterson peak value at a specific site $i$ can be written as

$$P(\vec{r}_i) \propto \alpha_{0i} \qquad (11)$$

Thus, $\alpha_{0i}$ at a specific site $i$ can be directly obtained from Patterson function value $P(\vec{r}_i)$ in a binary alloy.

The Fourier transform of 2D diffuse scattering pattern gives the projected Patterson function (**Fig. 3D,E**, the main text). The interpretation of these patterns can be made by approximating CrCoNi as a pseudo-binary alloy of $CrX_2$, with X = Co or Ni, because of the small difference between the scattering potential of Co and Ni. Using this approximation, we find $\alpha_{01} = -0.18$ and $\alpha_{02} = 0.38$ for the nearest and second nearest neighbor SRO parameter in the $L1_2$ CSRO, respectively. The estimated $\alpha_{01}$ is very close to the values of $\alpha_{CrCo}$ and $\alpha_{CrNi}$ reported by Tamm et al.(*8*) and Ding et al.(*9*).

Note 5. Guinier and Kratky analysis of small-angle neutron scattering

Guinier analysis allows model free determination of the radius of gyration ($R_g$) of a scattering object from small-angle X-ray or neutron scattering data (*12*). Guinier analysis is based on a series expansion of small-angle scattering intensity, leading to an expansion as follows

$$\ln[I(Q)] = \ln[I_0] - (R_g^2/3)Q^2 \qquad (12)$$

in which $Q$ is the scattering vector, $I(Q)$ is the measure small-angle neutron scattering (SANS) intensity, $I_0$ is the intensity when $Q$=0. $R_g$ is the radius of gyration represents the effective size of the scattering object. If the scattering objects have a sphere shape, the radius of the particles can be determined from

$$R_g^2 = \frac{3}{5}R^2 \qquad (13)$$

Linear (Guinier) scaling suggests that the system is essentially monodisperse.

Note 6. Chemical composition and compositional fluctuations

The STEM/EDX technique was employed to determine chemical homogeneity and chemical fluctuations in CoCrNi. To avoid the zone axis electron channeling effects (*13, 14*), we used the off-zone axis orientation (tilted ~2° away) to collect the EDS spectra over the areas of 100x100 nm$^2$ at a 1 nm step size. An electron probe of 1.2 nm in full-width-half maximum (FWHM) was used for the analysis. The sample thickness in both cases was determined as ~50 nm by CBED. The EDS spectra were acquired using a four-quadrant Super-X detector (Thermo-Fisher Scientific) with the acquisition time of 1s per spectrum. The beam current was kept at 150 pA for both samples. The K-α peaks in each X-ray spectrum were used to calculate the atomic percentage of each element. **Figure S13** shows the elemental chemical maps for Cr, Co and Ni in Samples WQ and HT, respectively. The mean compositions of Cr, Co and Ni are determined at 33.86 atomic percent (at%), 33.27 at%, and 32.87 at% in Sample WQ, and 33.50 at%, 33.89 at% and 32.62 at% in Sample HT, respectively. Both are close to the targeted composition.

Note 7. Calculation of radius distribution functions (RDFs) from APT

We performed the radial distribution function (RDF) in order to assess ordering/clustering within the data. The RDF is an average radial composition profile from every atom of interest (defined by the user) in the APT dataset (*15*). An APT dataset is a labelled point cloud with approximate atom positions defined by Cartesian coordinates. Any distribution function can be generated computationally within the dataset. This is not possible with conventional Fourier analysis of diffraction data. The Co and Ni centered RDFs are displayed in **Fig. S4** for the heat-treated and water quenched samples. There is a strong indication of Ni-Ni clustering and a weak indication of Co-Co clustering in both samples. There is also Cr depletion in proximity to the Ni and Co atoms. The presence of Co-Ni clustering in **Fig. S4A** and **S4D** and the absence of Ni-Co clustering in **Fig. S4B** and **S4E** is most likely due to Co being present within the Ni clusters and the stronger Ni-Ni clustering compared to Co-Co clustering. Our observation of Cr depletion around Ni and Co segregation is consistent with the finding of He bubble shells in irradiated NiCoCr alloys (*16*). In irradiated NiCoCr alloys, Ni and Co enrichment and Cr depletion around He bubbles is due to diffusion mechanisms, e.g. Ni and Co sites facilitate interstitial diffusion, while Cr sites facilitate vacancy diffusion.

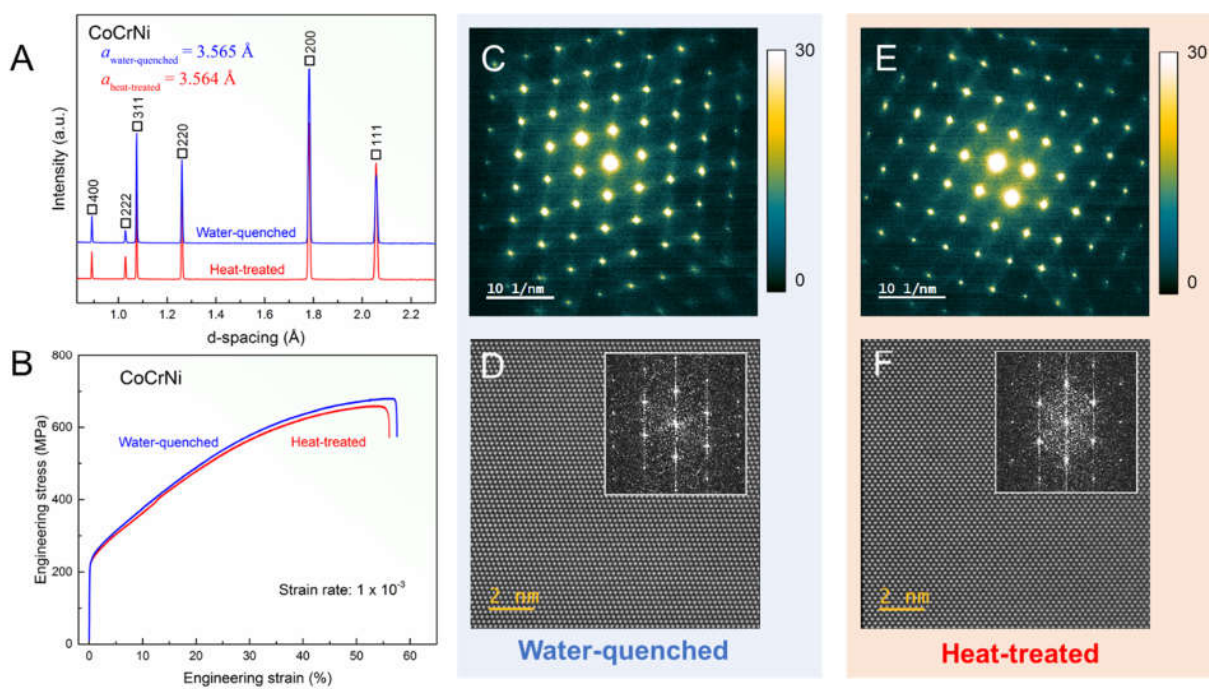

**Fig. S1. Characterization of the water-quenched and the heat-treated CrCoNi samples.** (**A**) Neutron diffraction patterns, showing the presence of the single fcc phase in both samples. (**B**) The measure tensile stress-strain up to fracture at the strain rate of $1 \times 10^{-3}$ s$^{-1}$. (**C, E**) Energy-filtered electron diffraction patterns and (**D, F**) atomic resolution STEM imaging. The insets in (D, F) are power spectra of the images.

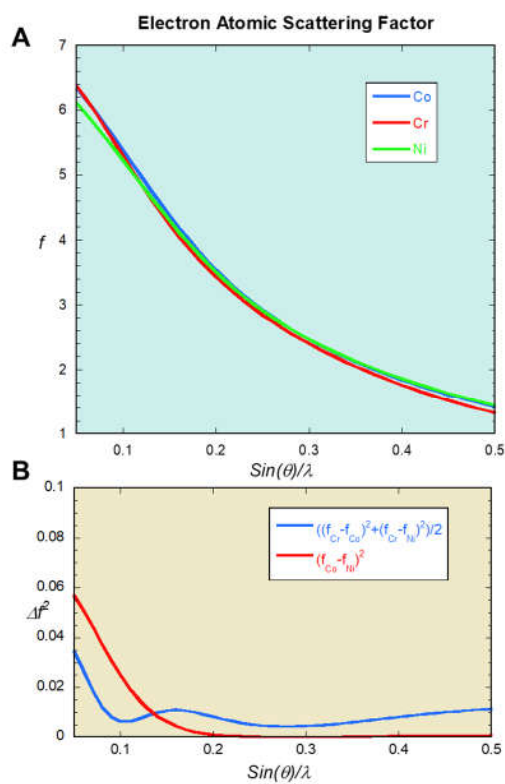

**Fig. S2.** (**A**) Atomic scattering factors of Co, Cr and Ni, and (**B**) their square differences.

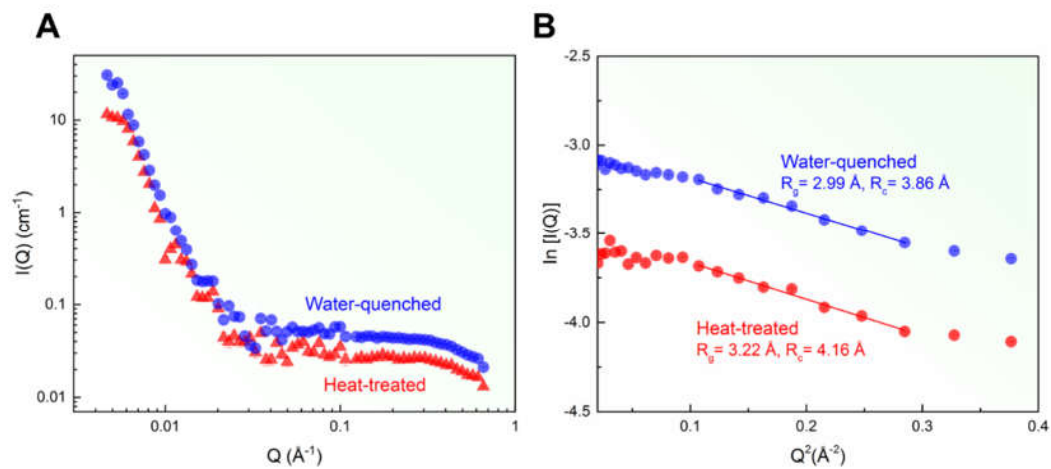

**Fig. S3.** (**A**) SANS-intensity distribution as a function of momentum transfer, Q, of the water-quenched and heat-treated CrCoNi samples. (**B**) Guinier plots plotting In[I(Q)] vs $Q^2$ and Guinier fit at a high Q range.

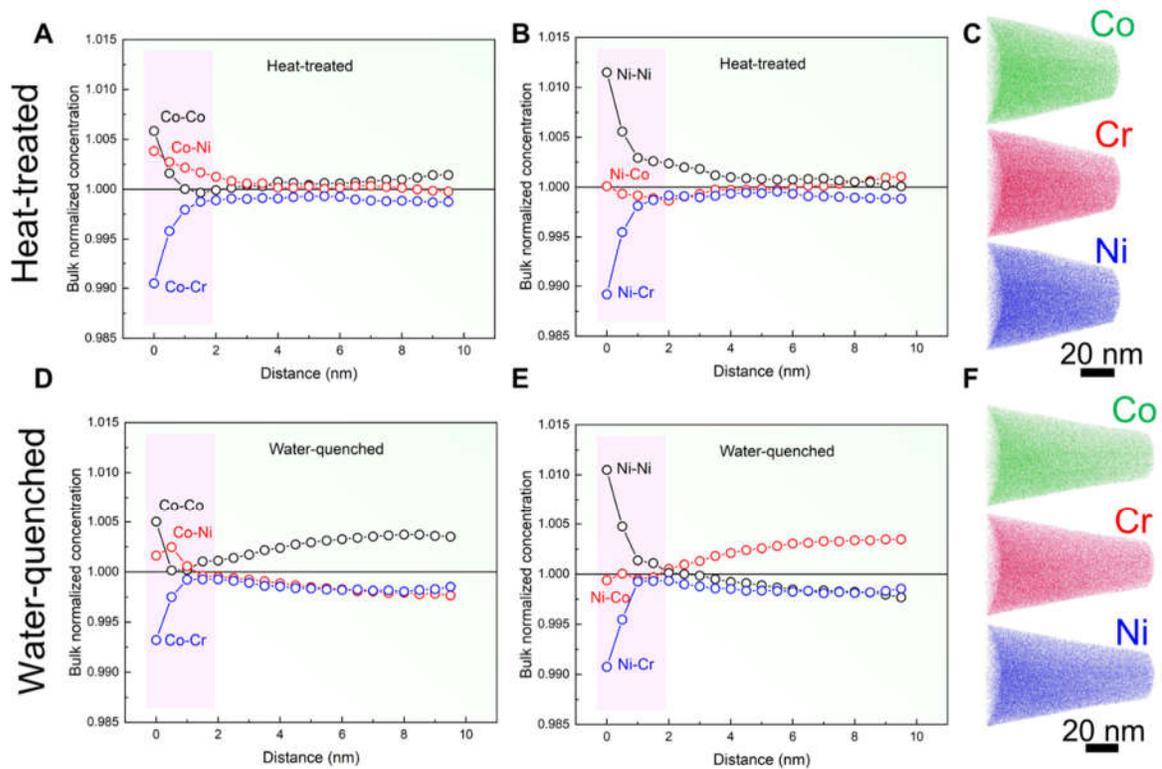

**Fig. S4. APT results of the heat-treated and water-quenched CoCrNi samples.** (**A**) and (**B**), and (**D**) and (**E**) calculated partial RDFs with Co and Ni as the center atom, respectively, as determined from APT data, for heat-treated and water-quenched CoCrNi samples, respectively. Note that the chemical concentrations were normalized for the bulk composition of the 25×25×25 nm cube selected for the RDF. (**C**) and (**F**) APT atom maps of the constituent elements (Co, Cr, and Ni) for heat-treated and water-quenched CoCrNi samples, respectively.

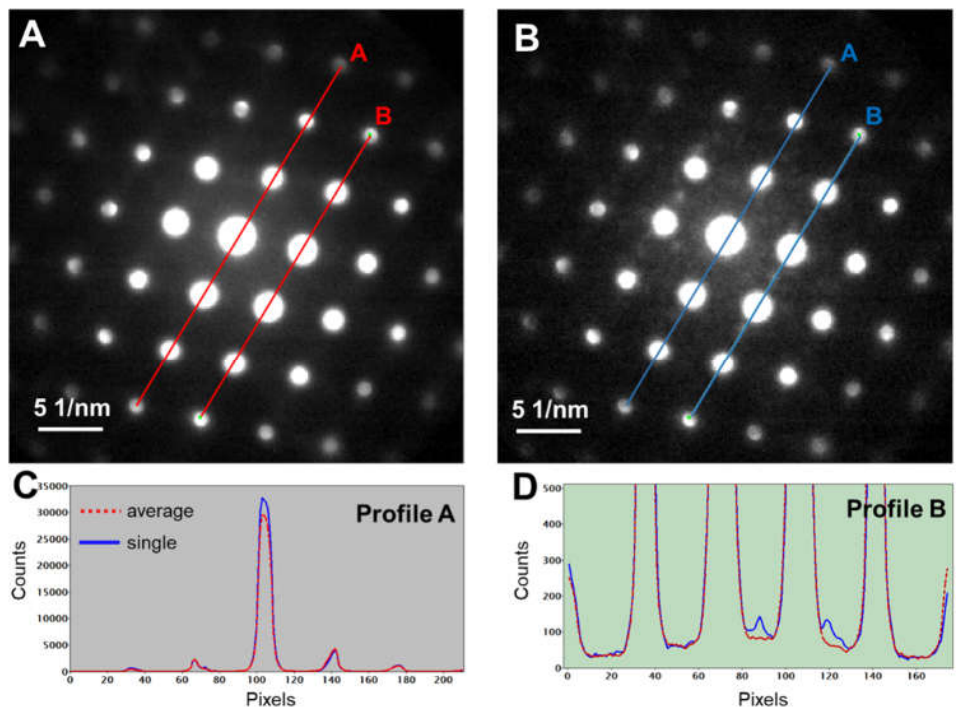

**Fig. S5. The collected DP dataset from Sample HT.** (**A**) the averaged DP from ten thousand DPs, (**B**) a single DP showing local diffuse scattering, (**C**) and (**D**) intensity profiles from lines A and B marked in (A) and (B). The DPs are along the [110] zone axis.

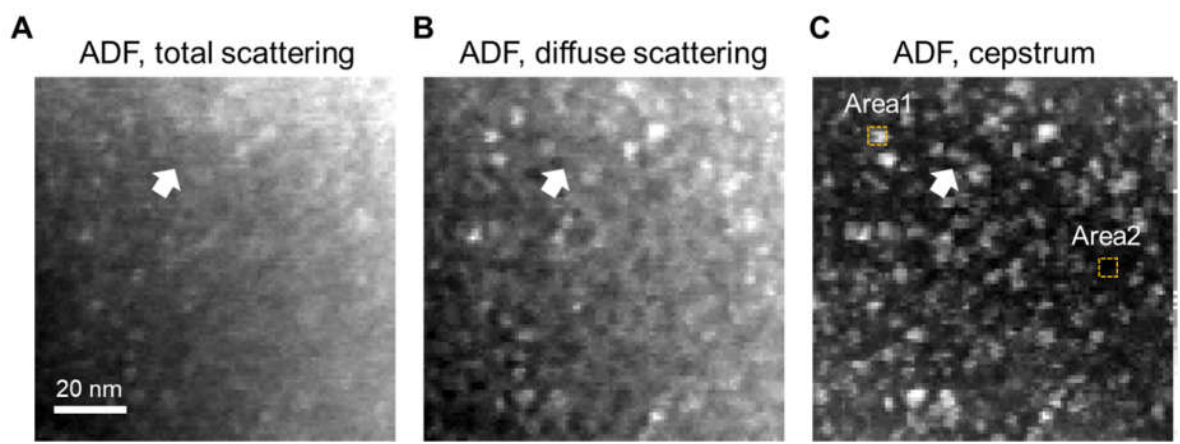

**Fig. S6. ADF images reconstructed from the 4D diffraction dataset.** For sample WQ, the reconstruction is carried out by using (**A**) total scattering intensity, (**B**) diffuse scattering intensity, and (**C**) diffuse scattering intensity after cepstral transformation. The arrows mark an example of a weak NC detection by the three imaging methods.

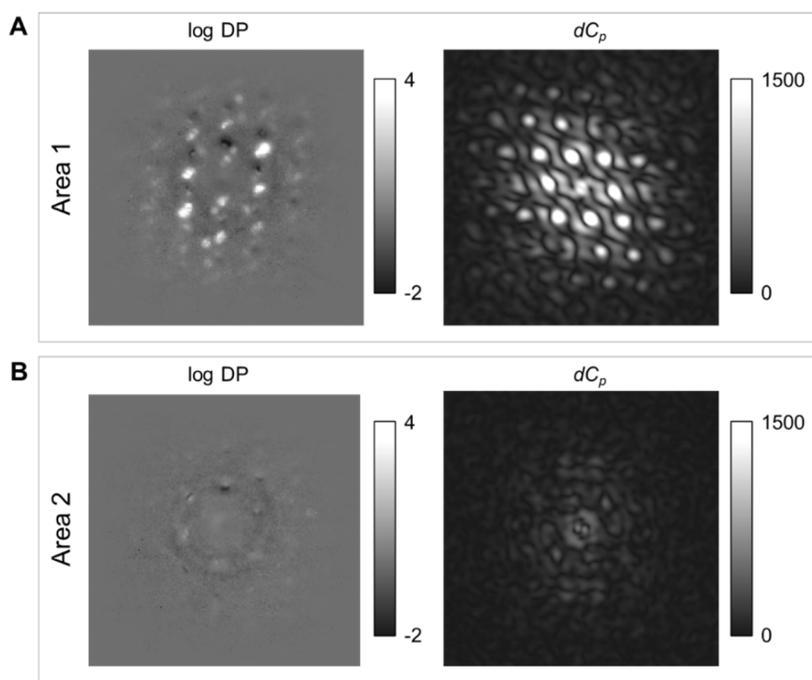

**Fig. S7.** Bragg reflection removal and difference cepstrum analysis for (**A**) Area1 and (**B**) Area2 in Suppl. Fig. 2c. Here log DP refers to log difference DP as given in equation S1.

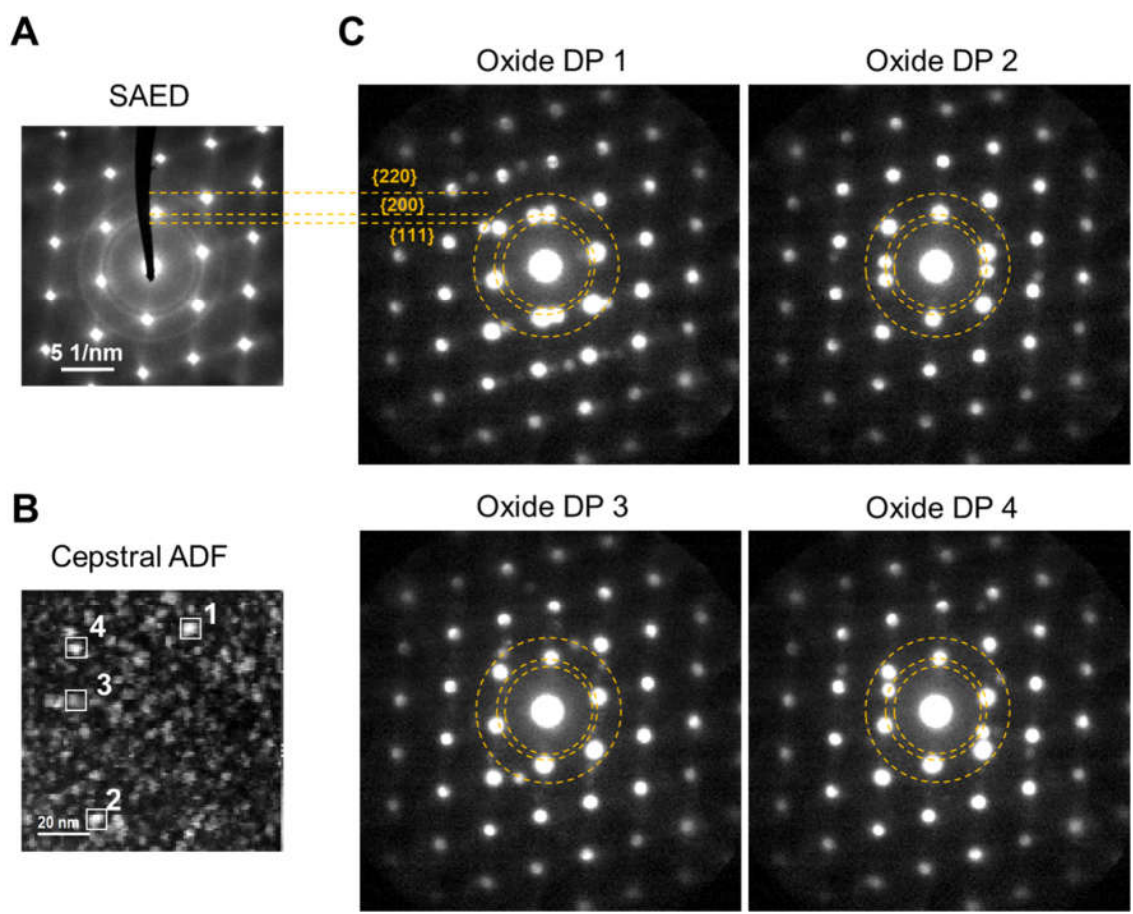

**Fig. S8. Identification of surface oxides in electron diffraction patterns.** (**A**) EF-SAED pattern recorded from Sample WQ after extended exposure to air showing the powder rings from surface oxides with the NiO structure. (**B**) A cepstral ADF image with marked surface oxide particles. (**C**) Oxide diffraction patterns corresponding to the four marked oxide particles in (B) with extra reflections falling on the rings.

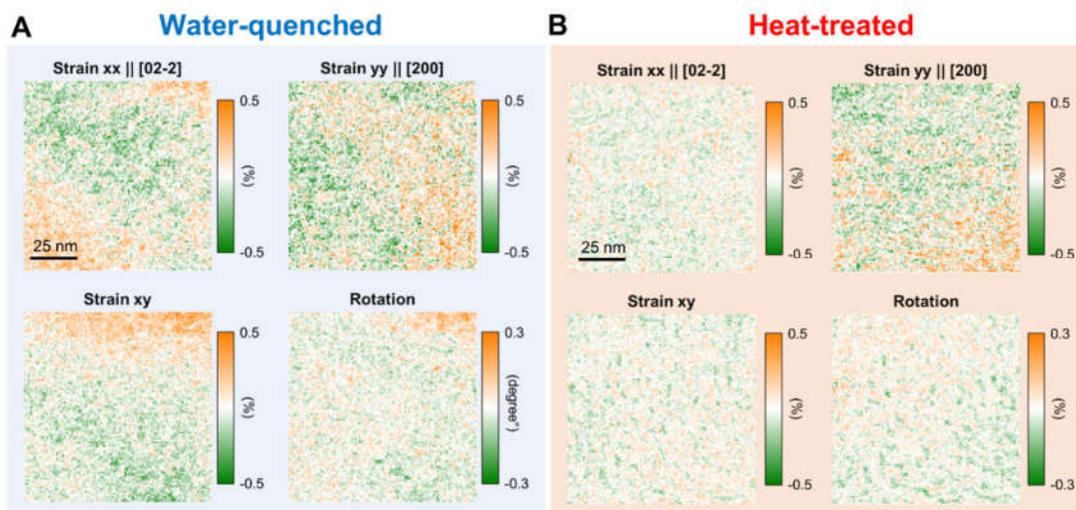

**Fig. S9.** The measured 2D strain with x || [02-2] and y || [200] directions projected along the [011] zone axis for (**A**) Sample WQ and (**B**) Sample HT.

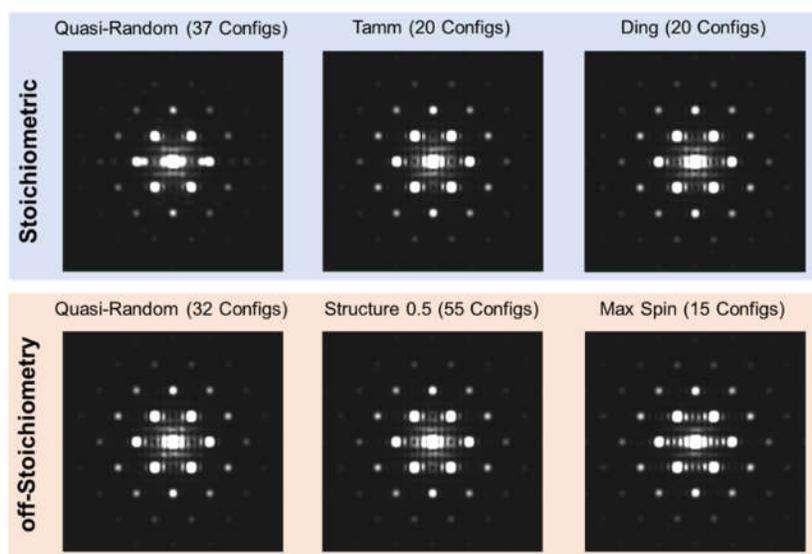

**Fig. S10.** Configurations averaged [110] zone axis diffraction patterns calculated with the atomistic models based on the WC SRO parameters for Tamm, Ding, and Structure 0.5, and zero SRO parameters for the quasi-random models (*7*). The Max Spin model optimizes atomic pairing for spin ordering (*7*). The number of atomistic configurations (configs) for each DP is indicated at top.

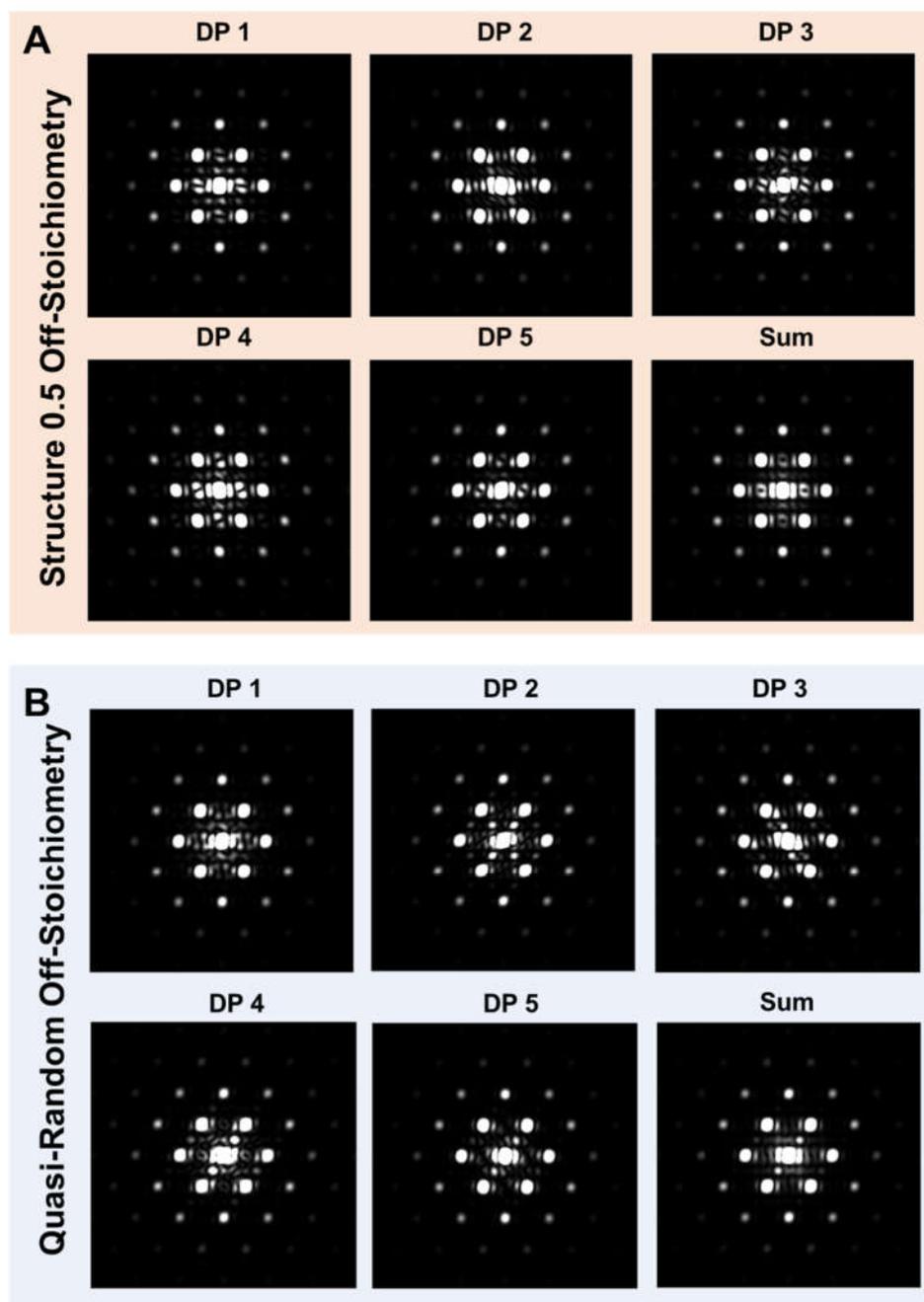

**Fig. S11.** The [110] zone axis diffraction patterns calculated with five configurations in the off-stoichiometry (**A**) Structure 0.5 and (**B**) quasi random atomistic models and their sum.

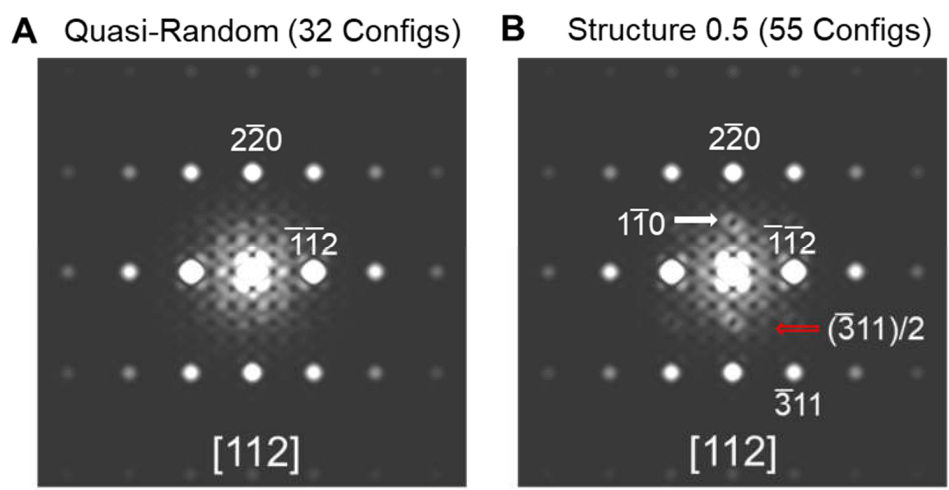

**Fig. S12.** Simulated electron diffraction patterns along the [112] zone axis for the off-stoichiometry quasi-random and structure 0.5 models.

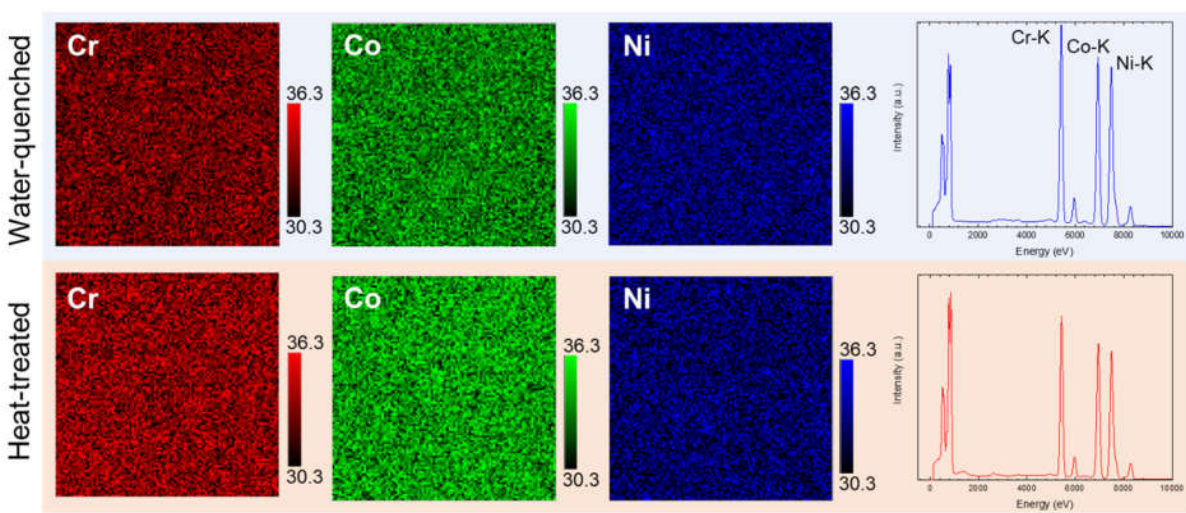

**Fig. S13.** The elemental chemical maps obtained and the average EDS spectra from the STEM-EDS datasets for the water-quenched and heat-treated samples.

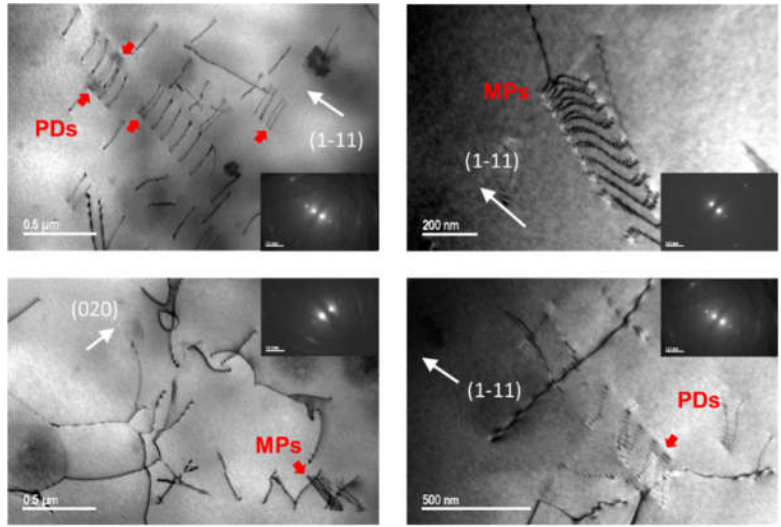

**Fig. S14.** Two-beam bright-field images of dislocations in the water-quenched CrCoNi sample, showing two additional dislocation configurations, paired dislocations (PDs) and dislocation multipoles (MPs), within the short slip trace, which are mixed with the wavy dislocation slips (Fig. 4, the main text) in the slightly deformed sample.

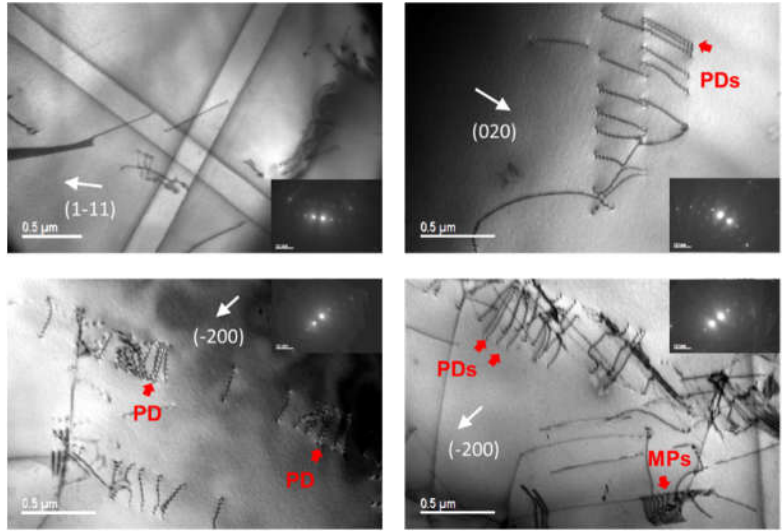

**Fig. S15.** Two-beam bright-field images of dislocations in the heat-treated CrCoNi sample, showing long slip traces as well two different dislocation configurations (paired dislocations (PDs) and dislocation multipoles (MPs)) within the slip traces in the slightly deformed sample.